\documentclass[twocolumn,prd,showpacs,10pt,superscriptaddress]{revtex4}
\usepackage{graphicx}

\def\AA{{\cal A}}
\def\BB{{\cal B}}

\def\LL{{\cal L}}

\def\etal{{\em et al.}}
\def\be{\begin{equation}}
\def\ee{\end{equation}}
\def\bea{\begin{eqnarray}}
\def\eea{\end{eqnarray}}
\def\alm{a_{\ell m}}
\def\smax{s_{\rm max}}
\def\lmax{\ell_{\rm max}}
\def\tr{{\rm tr}}

\def\id{1}

\def\Natur{Nature}

\newcommand{\UUNIT}[2]{\;{\mathrm{#1}^{#2}}}
\newcommand{\ddd}{{\rm d}}
\newcommand{\LSS}{{\rm LSS}}
\newcommand{\SCAL}{{\rm S}}

\begin{document}

\title{On the detectability of non-trivial topologies}

\date{\today}

\author{M.\ Kunz} 
\email{Martin.Kunz@physics.unige.ch}
\affiliation{D\'{e}partement de Physique Th\'{e}orique, Universit\'{e}
de Gen\`{e}ve, 24 quai Ernest Ansermet, CH-1211 Geneva 4, Switzerland}
\author{N.\ Aghanim} 
\email{Nabila.Aghanim@ias.u-psud.fr}
\affiliation{IAS, CNRS \& Univ. Paris-Sud, B\^{a}t. 121, F-91405,
Orsay Cedex, France} 
\author{A.\ Riazuelo} 
\email{riazuelo@iap.fr}
\affiliation{Institut d'Astrophysique de Paris, UMR7095 CNRS,
Universit\'e Pierre \& Marie Curie, 98 bis boulevard Arago, 75014 Paris,
France} 
\author{O.\ Forni} 
\email{Olivier.Forni@cesr.fr}
\affiliation{CESR, 9 Avenue du Colonel Roche, B.P.\ 4346, F-31028
Toulouse, Cedex 4, France}

\begin{abstract}

We explore the main physical processes which potentially affect the
topological signal in the Cosmic Microwave Background (CMB) for a
range of toroidal universes.  We consider specifically reionisation,
the integrated Sachs-Wolfe (ISW) effect, the size of the causal
horizon, topological defects and primordial gravitational waves. We
use three estimators: the information content, the S/N statistic and
the Bayesian evidence. While reionisation has nearly no effect on the
estimators, we show that taking into account the ISW strongly
decreases our ability to detect the topological signal. We also study
the impact of varying the relevant cosmological parameters within the
$2\sigma$ ranges allowed by present data. We find that only
$\Omega_\Lambda$, which influences both ISW and the size of the causal
horizon, significantly alters the detection for all three estimators
considered here.

\end{abstract}

%\keywords{cosmology: CMB anisotropies}
%\pacs{98.80.Es}
\maketitle

\section{Introduction}

One of the fundamental questions of cosmology is whether the universe
is infinite or finite, a question which is especially relevant in the
context of string theory where several dimensions may be compact. With
the move towards precision cosmology, it has become possible to study
the topology of the universe up to the causal horizon which
fundamentally limits our view of the world.

Many tests have recently been developed to detect and constrain the
topology of the universe, see Refs.~\cite{rep1,rep2} and references
therein.  Observationally a non-trivial topology can be detected
directly with the observed map of CMB temperature and polarisation
fluctuations through matched and correlated circles in the
CMB~\cite{circle,cornish2,roukema,luminet,cress,riap}. The topology
can also be searched for in the expansion of the CMB map in terms of
spherical harmonics, $T(x) = \sum_{\ell, m} \alm Y_{\ell m}(x)$, where
$x$ are the pixels and both the temperature field and the $\alm$ are
random variables. The presence of a non-trivial topology introduces
preferred directions, which breaks global isotropy and induces
correlations between off-diagonal elements of the two-point
correlation matrix $\left<\alm a_{\ell'm'}^*\right>$. Methods to find
such off-diagonal correlations and to put optimal constraints on the
size of the universe were proposed in Ref.~\cite{topo1}. They are
based on matching the measured correlations to a given correlation
matrix through a correlation coefficient, as well as a maximum
likelihood method and Bayesian model comparison.

So far the focus was purely on trying to detect any non-trivial
topologies. However, to determine reliable limits, it is also
necessary to understand all effects which can affect the detection.
For this purpose, we study the impact of changing the cosmological
parameters in the ranges allowed by present data. This should be seen
as a step towards a rigorous search for the presence of signs of
non-trivial topologies in the CMB. Eventually, such an analysis should
account for other contributions like foreground emissions as well as
sky cuts.

One additional reason to perform such a detailed analysis comes from
statistics. The potentially most direct way to quantify the
probability that our universe is infinite is to compute the model
probability with Bayesian evidence. To do this, one needs to integrate
over all parameters, which means that one needs to know the behaviour
of the likelihood for a topology under changes of the parameters.

This paper is organised as follows: we start by reviewing our notation
and the likelihood-based probes of topology. We then study the main
physical effects that are relevant for topology studies. Finally we
discuss how changes in the cosmological parameters affect our
detection ability, and what this means for realistic universes where
the parameters are not known with absolute certainty.

\section{Methodology}

In this section, all the simulations (as described in
Refs.~\cite{riaz1,riaz2}) are based on a flat $\Lambda$CDM model with
$\Omega_\Lambda = 0.7$, a reduced Hubble parameter of $h = 0.01 H_0
\times (1\UUNIT{km}{}\UUNIT{s}{-1}\UUNIT{Mpc}{-1})^{-1} = 0.67$, a
Harrison-Zel'dovich initial power spectrum ($n_\SCAL = 1$) and a baryon
density of $\Omega_b h^2 = 0.019$.  With this choice of cosmological
parameters we find a Hubble radius $c / H_0 \approx 4.8\UUNIT{Gpc}{}$,
while the radius of the particle horizon is $R_{\rm h} \approx
15.6\UUNIT{Gpc}{}$.  In this paper we limit ourselves to cubic tori,
denoted as T[X,X,X] where X is the size of the fundamental domain, in
units of the Hubble radius.  As an example, T[4,4,4] is a cubic torus
of size $(19.3\UUNIT{Gpc}{})^3$.  The volume of such a torus is
nearly half that of the observable universe.  The diameter of the
particle horizon is about $6.5 c / H_0$
(cf. section~\ref{sec:horizon}).  We will focus on three different
sizes, T[2,2,2], T[4,4,4] and T[6,6,6]. We combine the $\ell$ and $m$
indices of the spherical harmonic coefficients to a single index $s
\equiv \ell (\ell + 1) + m$ and mix both notations frequently.

In this paper we use the likelihood based methods of
Ref.~\cite{topo1}, which we review here quickly.  The CMB likelihood
is taken to be a multivariate Gaussian distribution with covariance
matrix $\BB_{ss'} = \langle a_s^* a_{s'}\rangle$ and zero mean,
$\langle a_s \rangle = 0$. This covariance matrix provides the full
statistical description of the CMB for a given topology. The
likelihood function then is \be \LL(\BB) \equiv p(\alm|\BB) \propto
\frac{1}{\sqrt{|\BB|}} \exp\left\{-\frac{1}{2} \sum_{s,s'} a^*_s
\BB^{-1}_{ss'} a_{s'} \right\} ,
\label{eq:like}
\ee where $|\BB|$ is the determinant of the matrix $\BB$.  Any further
model assumptions like the values of the cosmological parameters are
implicitly included in the choice of $\BB$.

Often it is easier to consider the logarithm of the likelihood,
$\ln\LL = -1/2(\ln |\BB|+\chi^2+\smax \ln(2\pi))$ where we have
defined 
\be 
\chi^2 = \sum_{s,s'} a^*_s \BB^{-1}_{ss'} a_{s'} .
\label{eq:chi2}
\ee
Writing $\langle a_s^* a_{s'}\rangle = \AA_{ss'}$ for the expectation
value of the two point function of the ``sky'' $\alm$, we can compute
the expectation value of the $\chi^2$, 
\be 
\langle \chi^2 \rangle = \sum_{ss'} \langle a_s^* a_{s'}\rangle
\BB_{ss'}^{-1} = \tr(\AA\BB^{-1}) , 
\ee 
where we have used the hermeticity of the correlation matrices.  For
the two special cases of an infinite universe and one for which
$\AA=\BB$ (ie. the measured $\alm$ are drawn from the template
covariance matrix $\BB$), we find 
\bea 
\langle \chi^2 \rangle_\infty & = & \sum_s C_s \BB_{ss}^{-1} \\ 
\langle \chi^2 \rangle_\BB & = & \tr(\id) = \smax 
\eea
As the $\alm$ are Gaussian random variables, we expect to find that
$\chi^2$ is distributed with a $\chi^2$-like distribution.  The
general expression for the variance is rather cumbersome, but for the
two special cases we find
\be
\sigma^2_\BB \equiv 
\langle (\chi^2)^2 \rangle_\BB - \langle \chi^2 \rangle^2_\BB = 2 \smax
\ee
and
\be
\sigma^2_\infty = 2 \sum_{ss'} C_s C_{s'} |\BB^{-1}_{ss'}|^2 .
\ee
To improve the stability of the algorithm against mis-estimates of the
power spectrum we whiten the $\alm$ by dividing out the power spectrum
\be
C_\ell = \frac{1}{2\ell+1} \sum_m |\alm|^2 ,
\ee
via the prescription 
\bea
\BB_{ss'}
 & \rightarrow & \frac{\BB_{ss'}}{\sqrt{C_s C_{s'}}} \label{eq:norm1} \\
\alm 
 & \rightarrow & \frac{\alm}{\sqrt{C_\ell}} \label{eq:norm2}
\eea
The power spectrum of the $\alm$ has to be estimated from the data,
which changes their distribution. We found in Ref.~\cite{topo1} that
this tends to increase the sensitivity of the estimators discussed
below.

\subsection{Information content}

As discussed in Ref.~\cite{topo1} it is possible to study directly the
information content in the CMB sky. For Gaussian fluctuations, the
infinite universe contains a maximum of information (maximal entropy),
while a finite (``small'') universe is less random. Mathematically
this is expressed through the non-zero correlations in the
off-diagonal part of the two-point function which make it easier to
predict the temperature distribution (at least in theory).

A commonly used measure of information content is given by the
Kullback-Leibler (KL) divergence between two probability distributions
$p$ and $q$,
\be
D_{KL}(p||q) \equiv \int p \log\left(\frac{p}{q}\right) .
\ee
This quantity measures the relative information entropy between the
two distributions. It is often used in information theory. For
example, if information distributed according to $q$ is transmitted
with a coding based on $p$ then $D_{KL}(p||q)$ describes how much
bandwidth is wasted because of the wrong coding (the base of the
logarithm is equivalent to a choice of units in which the KL
divergence is expressed). It is also possible to view the KL
divergence as a regularised entropy for continuous random variables,
as it satisfies the Gibbs inequality, $D_{KL}(p||q)\geq0$ with
equality only for $p=q$. Here we use the latter interpretation.

In our case we are dealing with Gaussian probability distributions
characterised by zero mean and covariance matrices $\AA$ and $\BB$,
for which the KL divergence can be evaluated directly, 
\be 
\int p(\AA) \log\frac{p(\AA)}{p(\BB)}
 = \frac{1}{2} \left[ \log |\BB| - \log |\AA|
                                 - \tr\left( 1-\BB^{-1}\AA\right)\right] .  
\ee 
Unfortunately the KL divergence is not symmetric in its arguments,
making its interpretation somewhat difficult. It is also not
rotationally invariant, so that it depends on the relative orientation
of the coordinate systems in which the correlation matrices are
expressed.  We will therefore only consider a special case where we
take $p(\BB)$ to be the distribution of an infinite universe. In this
case the KL divergence is the relative entropy to the topology
described by the correlation matrix $\AA$. This is both rotationally
invariant (as the infinite universe is isotropic) and has a
well-defined interpretation. It is also very easy to compute: For the
infinite universe $\BB=\BB^{-1}=\id$. The whitening enforces
$\tr(\AA)=\smax=\tr(\id)$, so that the relative entropy of $\BB$ is
just 
\be 
H(\AA) \equiv D_{KL}(\AA||\id) = -\frac{1}{2} \log |\AA| .
\ee 
This quantity describes how much less information is present in CMB
data of the (easier to predict) finite universe, as compared to the
information of the CMB of an infinite universe. This gives an {\em
upper limit} on the chance to detect a given topology.

\begin{figure}[ht]
\begin{center}
\includegraphics[width=86mm]{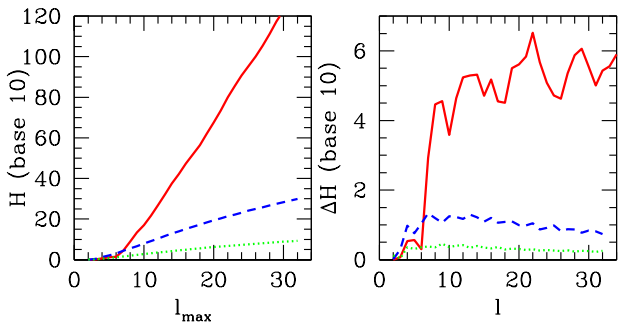}
\caption{\label{fig:info}
The information content as a function of the size of the fundamental
domain and of $\lmax$: The left panel shows $H$ and right panel the
increase per $\ell$, $\Delta H$, for (from the top downwards) 
T[2,2,2] (red solid line), T[4,4,4] (blue dashed line) and
T[6,6,6] (green dotted line).
}
\end{center}
\end{figure}

We plot in Fig.~\ref{fig:info} both information content $H$ as a
function of $\lmax$ (left panel) as well as the increase of $H$ per
$\ell$ (right panel) for the 3 topological models T[2,2,2] (red solid
line), T[4,4,4] (blue dashed line) and T[6,6,6] (green dotted line)
respectively. The information content decreases as expected for
topologies closer to an infinite universe (T[6,6,6] in our case).

\subsection{Measuring topology}
\label{sec:B}

While the previous subsection deals with a purely theoretical measure
of detectability, namely the information content in a given map or
matrix, we apply in this section two actual detection algorithms
discussed in Ref.~\cite{topo1}. For both we use likelihood,
Eq.~(\ref{eq:like}), or its logarithm $\chi^2$,
Eq.~(\ref{eq:chi2}). We exploit the fact that the the likelihood will
be higher if observed map and template agree.

\begin{figure}[ht]
\begin{center}
\includegraphics[width=70mm]{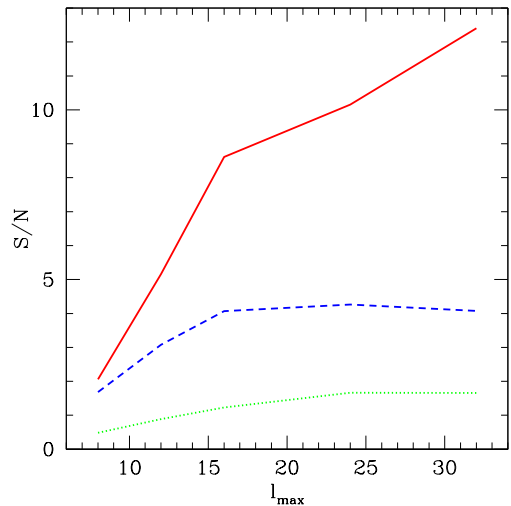}
\caption{\label{fig:ldm} S/N for (from the top downwards) T[2,2,2]
(red solid line), T[4,4,4] (blue dashed line) and T[6,6,6] (green
dotted line).}
\end{center}
\end{figure}

For the first analysis we compute the correlation between an observed
map and a template and we search the global minimum of $\chi^2$ over
all rotations of the $\alm$ (corresponding to relative orientations of
the observed map and template). In order to quantify possible false
detections of the topological signal, we give an estimate of the null
hypothesis, that is the possibility for an infinite universe to be
taken for a universe with a non trivial topology.  We use the
following procedure: For each given topology (T[2,2,2], T[4,4,4],
T[6,6,6]) we fix the template and generate several thousand random
maps both from the chosen template and for an infinite universe. We
then use the mean and standard deviation for both distributions to
compute a signal to noise statistic,
\be 
S/N(\BB,\chi^2)
 = \frac{|\langle \chi^2 \rangle_\infty - \langle \chi^2\rangle_\BB|}
        {\sqrt{\sigma(\chi^2)^2_\infty+\sigma(\chi^2)^2_\BB}} .
\label{eq:sn}
\ee 
This corresponds to the distance between the means in units of the
errors added in quadrature. We show results in Fig.~\ref{fig:ldm},
with the same colours/line styles as in Fig.~\ref{fig:info}. Again, we
see that larger universes are more difficult to detect. The T[6,6,6]
universe is barely detectable, while the others can be detect easily
for $\lmax>16$.

The above test is somewhat ad-hoc, so that we compute additionally the
probability that the $\alm$ are actually drawn from the distribution
described by the correlation matrix $\BB$. This is often called the
Bayesian evidence in astrophysical applications. Formally, the
probability of a model (specified here via $\BB$) is $p(\BB|\alm)$,
which we can connect to the likelihood, Eq.~(\ref{eq:like}) through
Bayes theorem, 
\be 
p(\BB|\alm) = p(\alm|\BB) \frac{p(\BB)}{p(\alm)} \propto \LL(\BB) p(\BB) .
\ee
We will be interested in ratios of the these probabilities for
different $\BB$ (describing different topologies). As we have no a
priori knowledge of the topology, we will take $p(\BB)$ to be
constant. However, the numerical value of $\LL(\BB)$ depends on the
orientation of the template, so that we need to marginalise over the
angles (see Ref.~\cite{topo1} for more details,
and~\cite{inoue,jaffe1,jaffe2} for other examples of the use of
Bayesian evidence in topology). Unfortunately the likelihood is very
strongly peaked in the vicinity of correct alignments. The numerical
integration is difficult, even using adaptive schemes.

\begin{figure}[ht]
\begin{center}
\includegraphics[width=70mm]{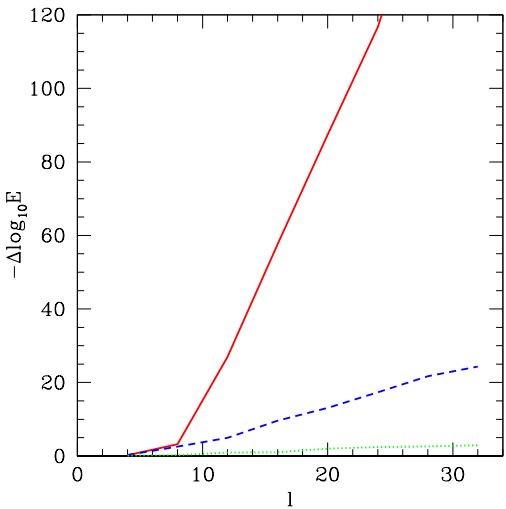}
\caption{\label{fig:ev}
Evidence for (from the top downwards) 
T[2,2,2] (red solid line), T[4,4,4] (blue dashed line) and
T[6,6,6] (green dotted line), compared to the $\alm$ of an
infinite universe.
}
\end{center}
\end{figure}

Having computed $p(\BB|\alm)$ we define a relative evidence for a topology
as the ratio of this probability to the one for an infinite universe,
\be
\Delta \log E (\BB) = \log p(\BB|\alm) - \log p(\id|\alm) .
\ee
This gives the relative probability that the $\alm$ are due to a universe
described by the topology and cosmology encoded in $\BB$, compared to
the probability that they are due to an infinite universe. Fig.~\ref{fig:ev}
shows the strong dependence of the evidence on the size of fundamental
domain, the figure looks somewhat similar to the one of $H$. One can further
eliminate the dependence on the cosmological parameters by marginalising
(integrating) over them as well. One of the motivations for this paper
is to study the dependence of $\Delta \log E (\BB)$ on the cosmological
parameters and to understand which integrals are necessary. Finally, by
also marginalising over a class of topologies (e.g.\ the X of cubic tori
T[X,X,X]) and using the actually measured $\alm$ we can compute the probability
that our universe has that kind of topology.

\section{Physical effects at large scales}

The shape of the universe leaves specific imprints on large scale CMB
anisotropies at multipoles lower than $\ell \sim 60$ that can be
detected via the correlation matrix of the $\alm$ coefficients. Such
scales suffer many problems. Firstly, it is at these scales that
cosmic variance $\Delta C_\ell =\sqrt{2/(2\ell+1)}C_\ell$ is the
largest. Then, at those scales several physical processes affect the
CMB signal which may change our ability to detect non-trivial
topologies and to measure their properties. In the following we will
study the main effects, namely the reionisation and the integrated
Sachs-Wolfe (ISW) effect, in addition to the relative size of the
particle horizon as well as tensor modes and defects. In the next
section, we will vary the relevant cosmological parameters and study
their influence on the detection of the topological signal.

\subsection{Relative size of fundamental domain and particle horizon}
\label{sec:horizon}

As discussed in the introduction, we fix the size of fundamental
domain in units of $c / H_0$. Another key scale is the distance to the
last scattering surface which is very close to the size of the
particle horizon. In the following, we will not discriminate between
these two. As light-rays move on geodesics with $\ddd s^2 = 0$ we can
compute the particle horizon today in comoving coordinates from

\be \eta_0 = \int_0^{t_0} \frac{c\;\ddd t}{a(t)} = c \int_0^1
\frac{\ddd a}{a^2 H(a)} .  \ee 

The quantity relevant to us is the diameter of the particle horizon
$D$ (the ``size of the causal domain''), which, in units of $c / H_0$
is 
\be D = \frac{2 a_0 H_0 \eta_0}{c} = 2 \int_0^1 \ddd x \left(
\omega_r h^{-2} + \Omega_m x + \Omega_\Lambda x^4 \right)^{-1/2} , \ee
where the new integration variable is the normalized scale factor $x
\equiv a / a_0$, and we have assumed flat spacelike sections and dark
energy in the form of a pure cosmological constant. The energy
density in radiation $\omega_r \equiv \Omega_r h^2$ is fixed very
precisely by the temperature of the CMB (assuming massless neutrini)
to $4.2\times 10^{-5}$. Its presence changes the value of the above
integral roughly by $\sqrt{\Omega_r}\approx 0.01$, i.e. the residual
influence of the Hubble constant is of the order of $1\%$ only.  Since
we assume that the universe is flat, we have $\Omega_m =
1-\Omega_r-\Omega_\Lambda$. In reality, the important scale is rather
the distance to the last scattering surface (LSS), so that we should
set the lower limit of integration to $x_\LSS\approx1/1090$.
This modifies the result by about 2\% but does not change the
conclusions.

The strength of the topological signal depends (among other things) on the
ratio of the two scales. A universe that is much smaller than the
causal domain leaves a strong imprint in the CMB, while one that is
much larger will be impossible to distinguish from a truly infinite
universe.  The value of $D$ corresponds roughly to the largest $X$
that we can hope to detect.  We show in Figures~\ref{fig:info},
\ref{fig:ldm} and~\ref{fig:ev} how the topology becomes easier to
detect as the fundamental domain becomes smaller (for a fixed size of
the particle horizon).

\begin{figure}[ht]
\begin{center}
\includegraphics[width=70mm]{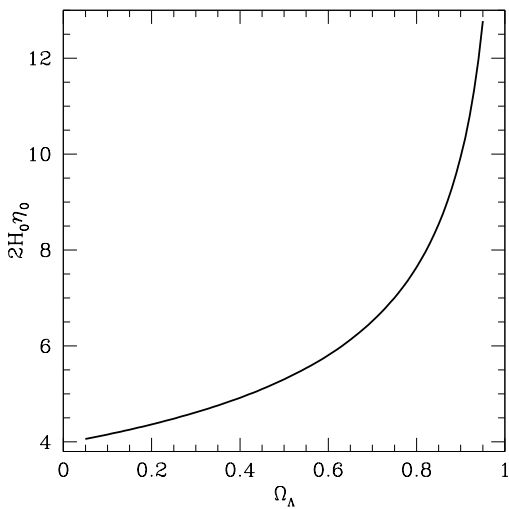}
\caption{\label{fig:D} Diameter of the particle horizon, $D = 2 a_0
\eta_0 H_0 / c$, for a flat universe and as a function of
$\Omega_\Lambda$. This corresponds roughly to the maximal fundamental
domain size $X$ that we can hope to detect.  }
\end{center}
\end{figure}

For a fixed size of the fundamental domain, the parameter affecting
the ratio of the two scales is $\Omega_\Lambda$ as it changes $D$, see
Fig.~\ref{fig:D}. If the matter density is very large, $\Omega_m
\approx 1$, then the size of causal domain corresponds roughly to the
size of a T[4,4,4] universe (in a purely matter-dominated universe,
the horizon size equals twice the Hubble radius, hence $D = 4$). On
the other hand, the causal domain of a flat low-density universe can
have an over 1000 times larger volume.

The importance of the correct orientation of the fundamental domain
was already discussed in Ref.~\cite{topo1}.  We found that the
orientation is one of the critical parameters and needs to be taken
into account in any analysis. This can be done by averaging quantities
over the orientation (as is done for a Bayesian study), or by taking
the maximum over all orientations (with the matched filter, or also
when looking for correlated circles).

\subsection{ISW effect}
\label{sec:isw}

The CMB photons are influenced by the change in the gravitational
potential through which they must pass. The effects arising from
time-variable metric perturbations are generally known as the
integrated Sachs-Wolfe effect~\cite{sacwol} in the linear regime and
go by the names of Rees-Sciama effect~\cite{rees86} and the effect
from moving halos~\cite{aghanim98} in the non-linear regime. The ISW
effect depends on the time derivative of the gravitational
potential. The anisotropies due to the ISW effect thus depend on the
parameters of the background cosmology and are also tightly coupled to
the clustering and the spatial and temporal evolution of the
intervening structures.  The most important contribution for
topological studies is the late ISW effect due to the decay of the
gravitational potential~\cite{KoSt85,KaSp94}. This happens in a low
matter density universe at the onset of dark energy (or spatial
curvature) domination, when the increased rate of expansion of the
universe reduces the amplitude of gravitational potential.

The power spectrum of the ISW can be written as a function of the
power spectrum of the potential $P_\Phi(k)$ like
\begin{equation}
C_\ell = (4 \pi)^2\ \int P_\Phi(k) k^2 \;\ddd k 
                        \left[\int_{\eta_\LSS}^{\eta_0}
                              2\dot{F}(\eta) j_\ell(kr)\;\ddd\eta \right]^2 \, ,
\label{eqn:Cl_ISW1}
\end{equation}
where the $\LSS$ subscript stands for the epoch of last scattering,
$j_\ell(x)$ is the spherical Bessel function and $r$ is the comoving
distance. $F(k,\eta)=D / x$ is the growth rate of the potential, where
$x$ is the (normalized) scale factor and $D$ is the linear growth factor. The
potential is related to the matter density contrast $\delta$ via the
Poisson equation, $k^2 \Phi = -\frac{3}{2} a_0^2 \Omega_m H_0^2 \delta
/ x$. In the matter-dominated era, the density contrast grows as $x$
and therefore the gravitational potential is constant. This is no
longer the case in the dark-energy dominated era.

The ISW effect is seen mainly in the lowest $\ell$-values of the power
spectrum where the cosmic variance is large. This makes its direct
measurement very difficult. Since the time evolution of the potential
that gives rise to the ISW effect may also be probed by observations
of the large scale structure, one expects the ISW to be correlated
with tracers of the large scale structure (see, e.g.\
Ref.~\cite{CrTu96}). This could conceivably help to remove it at least
partially.
\begin{figure}[ht]
\begin{center}
\includegraphics[width=70mm]{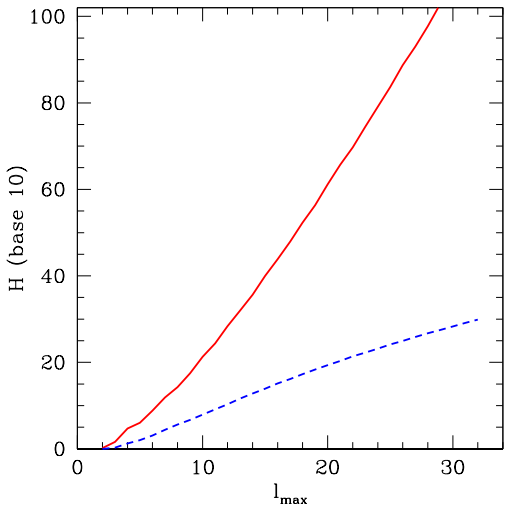}
\caption{\label{fig:isw_h}
Influence of the ISW effect on the information content: The
dashed blue line shows the $H$ for the T[4,4,4] model
while the red line shows the information content for the same
model but without ISW effect. The difference is very important.
}
\end{center}
\end{figure}
We show in Fig.~\ref{fig:isw_h} the information content $H$ of our
standard T[4,4,4] model (dashed blue line) and compare it to a
fictitious case where the ISW effect was turned off (red solid line;
the anisotropies are then only generated at last scattering). The ISW
is clearly a very strong effect that needs to be taken into account
correctly. We also checked its influence on the detection algorithms
discussed in section~\ref{sec:B} and found the same result.

\subsection{Reionisation}
\label{sec:reion}

At reionisation, CMB photons scatter off free electrons, altering the
CMB temperature anisotropies and inducing a polarised signal.  The
relevant effects in the case of topological studies are mostly those
induced by reionisation on the CMB temperature anisotropies. At
reionisation, the photon-electron interactions randomise the
directions of propagation of the fraction of rescattered photons. As a
result, the primary CMB anisotropies are suppressed and the power in
the acoustic peaks is damped by a factor $\exp{(-2\tau)}$, where
$\tau$ is the Thomson optical depth. Scales approaching the horizon at
reionisation are barely affected by the scattering, the suppression is
essentially efficient at small scales.

\begin{figure}[ht]
\begin{center}
\includegraphics[width=70mm]{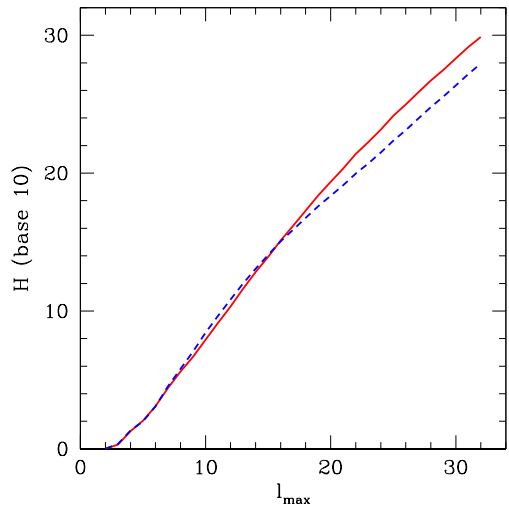}
\caption{\label{fig:reion_h}
Influence of reionisation on the information content: The
dashed blue line shows the $H$ for the T[4,4,4] model
while the red line shows the information content for the same
model but without reionisation. The difference is negligible.
}
\end{center}
\end{figure}

We investigate the effect of reionisation on the detectability of a
non-trivial topology in Fig.~\ref{fig:reion_h} by comparing a universe
with no reionisation (red solid line) and a universe with an optical
depth compatible with the WMAP1 constraints ($\tau = 0.166$, blue
dashed line). The induced changes are negligible. The latest WMAP
results suggest a significantly lower optical depth, therefore, the
effect of reionisation is even smaller.

\subsection{Tensor modes and defects}

Both gravitational waves and defects induce perturbations not
primarily at decoupling, but throughout the whole evolution of the
universe. Much of their contribution to the temperature anisotropies
is therefore due to their late-time behaviour. In this they 
behave similarly to the integrated Sachs-Wolfe effect, and their
contribution to the $\alm$ will be mostly uncorrelated with the
topological signal.

In a topologically non-trivial universe defects can stretch across the
full fundamental domain (if they are at least one-dimensional). This
may lead to a preferential alignment in certain directions. It also
means that they cannot be larger than the size of the fundamental
domain, so that their scaling behaviour has a cut-off at that
scale. When the universe becomes larger, the defects will generically
disappear (up to a possible net winding charge), leading to less
fluctuation power on large scales.

Analyses~\cite{wmap3_spergel,us} have restricted the contributions of
primordial tensor modes and defects to be subdominant, and future
measurements will improve limits substantially. We will consider them
as a negligible source of ``noise'' for measurements of the
topology. If ever their presence is detected, we should take them into
account explicitely.

\section{Sensitivity to cosmological parameters}

We choose a fiducial model motivated by the WMAP3+SNLS results for a
flat $\Lambda$CDM model~\cite{wmap3_spergel}, with $H_0 =
72.4\UUNIT{km}{}\UUNIT{s}{-1}\UUNIT{Mpc}{-1}$, $n_\SCAL = 0.95$,
$\Omega_\Lambda = 0.75$, $\Omega_b h^2 = 0.0234$, and $\tau = 0.085$.
We then focus on the most relevant cosmological parameters and we
explore their effects on the detection of the topological signal by
varying in turn $H_0$, $n_\SCAL$ and $\Omega_\Lambda$ (keeping all
other parameters fixed). Table~\ref{tab:param} shows the range of
variation, corresponding to roughly the two-sigma limits of WMAP3
(with the exception of $\Omega_\Lambda$ where we also investigate
other values). We do not expect the normalisation $A_\SCAL$ or the
baryon density $\Omega_b h^2$ to play an important role. In
section~\ref{sec:reion} we further saw that there is only a small
difference between $\tau = 0$ and the fiducial case. We conclude that
variations of $\tau$ are also uncritical for measuring topology.

\begin{figure}[ht]
\begin{center}
\includegraphics[width=80mm]{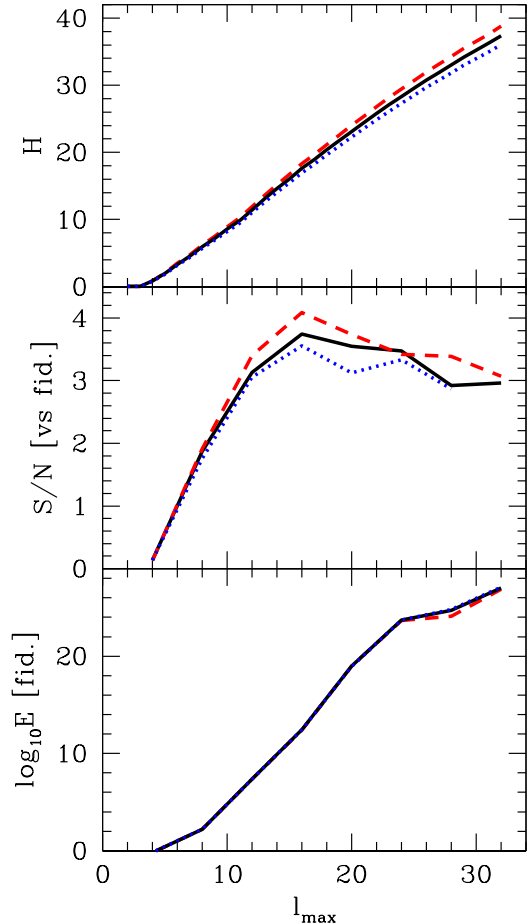}
\caption{For T[4,4,4], the effect of varying $n_\SCAL$: The top panel
shows the information content, for the fiducial model (solid black
line, $n_\SCAL = 0.950$) as well as $n_\SCAL = 0.916$ (dashed red line)
and $n_\SCAL = 0.982$ (dotted blue line). The middle panel (S/N) and
bottom panel (evidence) show how the detection changes when testing
simulated maps against the fiducial model. The black solid line uses a
map derived from the fiducial model, while the red dashed and blue
dotted lines are for the smaller and larger $n_\SCAL$ respectively.  }
\label{fig:ns_p}
\end{center}
\end{figure}

We limit ourselves to showing the results for the T[4,4,4] case since
it is more realistic than T[2,2,2] while still giving a clear signal
(as opposed to T[6,6,6]). The basic effects due to the parameter
uncertainties do not depend on the size of the fundamental domain.
\begin{table}
\begin{tabular}{rl}
$h =$ & $\{ 0.678, 0.724, 0.770 \}$ \\
$n_\SCAL =$ & $\{ 0.916, 0.95.0, 0.982 \}$\\
$\Omega_\Lambda =$ & $\{ 0.50, 0.70, 0.75, 0.80, 0.90\}$ \\
\end{tabular}
\caption{Parameter values investigated.}
\label{tab:param}
\end{table}
We show in the top panel of Fig.~\ref{fig:ns_p} how the information
content $H$ depends on the three values of $n_\SCAL$ (cf
table~\ref{tab:param}).  The other two panels show how such a
variation of $n_\SCAL$ affects the detection of a T[4,4,4] topology based
on our fiducial model. To this end we simulated maps with the
different values of $n_\SCAL$ (and all other parameters fixed to the
fiducial value) and then tested them against our fiducial model,
applying the algorithms of section~\ref{sec:B}. We find that the
information content does not vary significantly. Also the detections
measured by S/N (middle panel) and the Bayesian evidence (bottom
panel) do not change for $\lmax < 10$ with only small changes at higher
resolutions (within the statistical fluctuations induced by the map
realisations). We conclude that $n_\SCAL$ can safely be fixed to the
central value in table~\ref{tab:param}.

\begin{figure}[ht]
\begin{center}
\includegraphics[width=80mm]{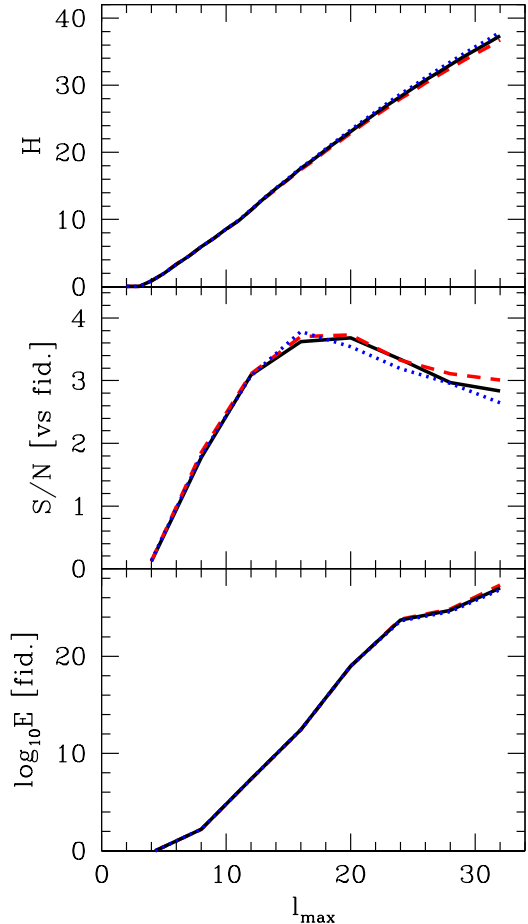}
\caption{The effect of varying $h$ for T[4,4,4]: The top panel shows
the information content, for the fiducial model (solid black line,
$h = 0.724$) as well as $h = 0.678$ (dashed red line) and $h = 0.770$
(dotted blue line). The middle panel (S/N) and bottom panel (evidence)
show how the detection changes when testing simulated maps against the
fiducial model. The black solid line uses a map derived from the
fiducial model, while the red dashed and blue dotted lines are for the
smaller and larger $h$ respectively.  }
\label{fig:h0_p}
\end{center}
\end{figure}

The reduced Hubble parameter $h$ intervenes both in the size of the
particle horizon (section~\ref{sec:horizon}) and in the ISW effect
(section~\ref{sec:isw}). We therefore expect some dependence on its
value. Repeating the same analyses as for $n_\SCAL$.  We find that the
results depend even less on $h$ than on $n_\SCAL$, see
Fig.~\ref{fig:h0_p}. Clearly, $h$ can also be fixed to the central
value in table~\ref{tab:param}.

\begin{figure}[ht]
\begin{center}
\includegraphics[width=80mm]{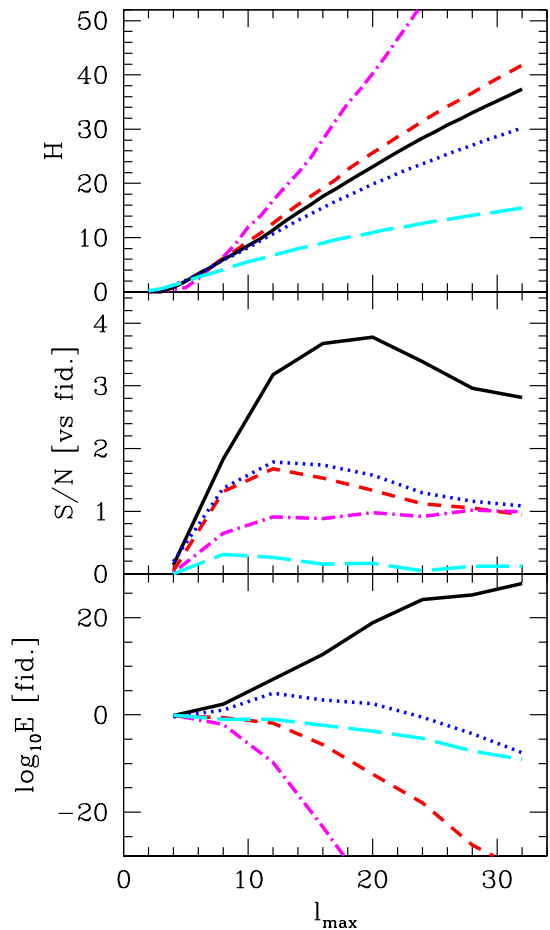}
\caption{The effect of varying $\Omega_\Lambda$ for T[4,4,4]: The top
panel shows the information content, for the fiducial model (solid
black line, $\Omega_\Lambda = 0.75$) as well as $\Omega_\Lambda =
0.80$ (dashed red line), $\Omega_\Lambda = 0.90$ (dot-dashed magenta
line), $\Omega_\Lambda=0.7$ (dotted blue line), and $\Omega_\Lambda =
0.50$ (long-dashed cyan line). The middle panel (S/N) and bottom panel
(evidence) show how the detection changes when testing simulated maps
against the fiducial model. The black solid line uses a map derived
from the fiducial model. The line styles of the other models are the
same as those of the information content (top panel).  }
\label{fig:ol_p}
\end{center}
\end{figure}

We follow the same procedure as above, varying $\Omega_\Lambda$ only,
for the values given in table~\ref{tab:param}. In the top panel of
Fig.~\ref{fig:ol_p} we plot the information content for these models,
with the largest $\Omega_\Lambda$ leading to the highest $H$. The
larger $\Omega_\Lambda$, the larger the causal domain, but also the
bigger the ISW contribution. The graph suggests that the effect due
to the size of the fundamental domain dominates over the increase in
``ISW noise''. As before, we then investigate how the detection changes
if map and template have a different $\Omega_\Lambda$. In contrast to
the two previous parameters, we find that there is a strong variation
in the detectability. We only have a high S/N if map and template agree
(black solid line in the middle panel). The maps with the next higher 
(red dashed line) and lower (blue dotted line) values of $\Omega_\Lambda$
are already below a significant detection limit, and the other cases
lie even lower. The evidence shows the same behaviour, with the fiducial
template leading to a much higher model probability than the others.
A full analysis requires to explore all acceptable values of the
cosmological parameters. Given that the topological signal remains
nearly constant for all parameters except $\Omega_\Lambda$, it
suffices to vary only $\Omega_\Lambda$ within the $2 \sigma$ region
allowed by WMAP3+SNLS.

\section{Conclusions}

We have studied cubic torus universes, investigating the change of the
topological signal with respect to the following effects: ISW,
reionisation, size of causal horizon $D$, topological defects and
primordial gravitational waves.  We consider these to be the
potentially most relevant effects for studying topology. We find that
reionisation has negligible effect.  The ratio $D$ to the size of the
fundamental domain is the key quantity which limits our ability to
detect a given topology.  The ISW effect in turn is not correlated
with the topological signal and so acts as an additional noise. The
same is true for topological defects, and tensor modes, but they are
experimentally constrained to a lower amplitude than the ISW
effect. We conclude that only the ISW effect and the size of the
causal horizon have to be considered.

An additional important parameter is the relative orientation of the
fundamental domain~\cite{topo1}. The methods we use take this already
into account so that we do not discuss it in detail.

When searching for traces of a non-trivial topology, we have to test a
priori templates for all cosmological models, varying the full set of
cosmological parameters. We have shown here that, given the
constraints from WMAP3+SNLS, only the allowed variation of
$\Omega_\Lambda$ is able to affect significantly the detections. This
parameter alters both the magnitude of the ISW effect and the size of
the causal horizon.  We find that the latter effect is more important
and templates for several values of $\Omega_\Lambda$ are required for
a complete analysis.

\begin{acknowledgments}
MK thanks Ruth Durrer for interesting discussions and acknowledges
financial support from the Swiss NSF. The authors acknowledge partial
support from Programme National de Cosmologie. NA and OF thank the
University of Geneva and MK thanks IAS for hospitality. We acknowledge
the use of HEALPix \cite{healpix} to manipulate the CMB maps.
\end{acknowledgments}

\end{document}